\documentclass[11pt,a4paper]{article}
\usepackage{jheppub}
\usepackage{tikz}
\usetikzlibrary{positioning}
\usepackage{hyperref}

\title{Holographic flavour and neural networks}

\author{Veselin G. Filev}
\affiliation{Institute of Mathematics and Informatics, Bulgarian Academy of Sciences\\
Acad. G. Bonchev Str., 1113 Sofia, Bulgaria.}
\emailAdd{vfilev@math.bas.bg} 
\date{} \abstract{In holography, flavour probe branes are used to introduce fundamental matter to the AdS/CFT correspondence. At a technical level, the probes are described by extremizing the DBI action and solving the Euler–Lagrange equations of motion. I report on applications of artificial neural networks that allow direct minimization of the regularized DBI action (interpreted as a free energy) without the need to derive and solve the equations of motion. I consider, as examples, magnetic catalysis of chiral symmetry breaking and the meson melting phase transition in the D3/D7 holographic set-up. Finally, I provide a framework which allows the simultaneous learning of the embeddings and the relevant aspects of the dual geometry based on field theory data.}

\begin{document}
\maketitle
\section{Introduction}

In recent years the AdS/CFT correspondence \cite{Maldacena:1997re} has developed into a powerful framework for studying strongly–-coupled gauge theories via dual gravitational descriptions. Of particular importance was the extension of the correspondence to gauge theories with quenched fundamental matter by introducing probe D$p$–branes whose dynamics are governed by the Dirac–Born–Infeld (DBI) action \cite{Karch:2002sh}.  In the probe limit, where ($N_f\ll N_c$), the backreaction of the flavour branes on the geometry can be neglected, which corresponds to quenched approximation in gauge theory. At a technical level this involves a choice of an ansatz for the brane embedding and solving the Euler–Lagrange equations of motion obtained by extremizing the DBI action (possibly supplemented by Wess–Zumino terms). Often this requires a detailed analysis of the numerical boundary–value problem and refinement of the required shooting methods.

Over the years these techniques have been applied to a variety of phenomena in strongly coupled gauge theories.  In this work we will focus on two important examples: magnetic catalysis of chiral symmetry breaking in the D3/D7 system \cite{Filev:2007gb} and the meson melting phase transition at finite temperature \cite{Babington:2003vm,Mateos:2006nu,Albash:2006ew}.  In all of these examples one must derive and solve non–linear ordinary differential equations to determine the brane profiles and compute thermodynamic quantities such as the free energy or condensate.

In parallel, the last few years have seen a surge of interest in applying machine learning techniques to problems in holography.  In a pioneering work, ref.~\cite{Hashimoto:2018ftp} showed how deep neural networks can be used to reconstruct parts of the bulk geometry from boundary correlation data.  Subsequent studies have explored reconstructing black–hole metrics from entanglement entropy or transport coefficients \cite{Ahn:2024jkk,Ahn:2024gjf,Lei:2025loy,Halverson:2024axc,Hashimoto:2024aga,Mansouri:2024uwc,Chen:2024ckb,Kou:2025qsg}, emergent spacetime from tensor networks \cite{Sahay:2024vfw,Akers:2024wre}, and the use of transformer architectures in holography \cite{Park:2023slm}.  Concurrently, physics–informed neural networks (PINNs) have been developed in the applied mathematics community to solve partial and ordinary differential equations by embedding the underlying variational or differential operators directly into the training loss.

In this work we bring physics–informed learning to holographic flavour physics by constructing artificial neural networks (ANNs) that directly minimize the regularized DBI action—interpreted as a thermodynamic free energy—without ever deriving or solving the Euler–Lagrange equations.  Concretely, we parameterize the D7–brane embedding $L(\rho)$ (or more generally $L(\rho,m)$ for a family of bare masses) by a feed–forward network and enforce the ultraviolet boundary conditions by a simple algebraic mapping.  The DBI free energy functional, suitably regularized by subtracting a reference embedding, becomes the differentiable loss function.  Automatic differentiation (via PyTorch \cite{Paszke:2019}) then yields the variation of the action with respect to the network parameters, and standard optimizers (e.g. Adam \cite{Kingma:2014adam}) find the equilibrium brane profile.

We illustrate this approach in two classic setups: magnetic catalysis of chiral symmetry breaking at zero temperature \cite{Filev:2007gb} and the meson–melting transition at finite temperature \cite{Babington:2003vm,Mateos:2006nu,Albash:2006ew}.  In both cases the ANN reproduces the known brane embeddings and thermodynamic curves (free energy versus mass, condensate versus mass) with high accuracy.  We then extend the method to learn the entire one–parameter family $L(\rho,m)$ with a single network, enabling efficient differentiation with respect to the bare mass.  Finally, inspired by the inverse–geometry program \cite{Hashimoto:2018ftp}, we tackle the inverse problem: given free energy data as a function of mass, we jointly train two networks—one for the embedding and one for an unknown radial potential $V(r)$ encoding the metric functions—via an alternating optimization scheme.  This recovers both the D7–brane profiles and the underlying bulk geometry (e.g.\ the AdS–Schwarzschild black–hole metric) purely from boundary data.

The rest of the paper is organized as follows.  In section 2 we introduce our ANN ansatz for extremizing the DBI action and present results for magnetic catalysis.  Section 3 applies the method to the meson–melting transition at finite temperature.  In section 4 we generalize to a conditional network learning $L(\rho,m)$ for all masses simultaneously.  Section 5 addresses the inverse problem of reconstructing the bulk geometry from free energy data.  We conclude in section 6 with a discussion of future directions.

\section{Extremising the DBI action with neural network}
One of the ways to introduce fundamental degrees of freedom in the context of holography is to consider probe D-branes \cite{Karch:2002sh}. In this paper we will focus on examples of D7--brane embeddings even though the approach that we are describing is directly applicable to any of the Dp/Dq intersections.  The DBI action is given by:
\begin{equation}\label{DBI}
S_{DBI} = {\mu_7}\int d^8\xi\, e^{-\Phi}\left[-{\rm det}\left(G_{ab} + B_{ab} + (2\pi\alpha')F_{ab}\right)\right]^{1/2}\ ,
\end{equation}
where $\mu_7 = [(2\pi)^7\alpha'^4]^{-1}$ is the D7-brane tension, $G_{ab}$ and $B_{ab}$ are the pull backs of the metric and the Kalb-Rammond $B$-field and $F_{ab}$ is the world-volume gauge field.  In most applications one considers an ansatz for the D7-brane embedding and solves the Euler-Lagrange equations which extremise the action (\ref{DBI}). In this work we will describe a method where after fixing the ansatz, we regularize the action (\ref{DBI}), with the results being interpreted as the thermodynamic free energy of the system and then use neural network to parameterize the embedding and train the network using the regularized free energy as the loss function during the training process.  

To formulate a finite free energy functional one could use the elegant scheme of ref. ~\cite{Karch:2005ms}, where an appropriate covariant counter term is added to the action to cancel the ``UV divergences'',  or equivalently one can use a subtraction regularization scheme where the free energy of a benchmark embedding is subtracted to keep the free energy finite. For simplicity in our examples we will use the latter approach.  We begin with the example of magnetic catalysis of chiral symmetry breaking in the D3/D7 system. 

\subsection{Learning in holographic magnetic catalysis}
We will consider AdS$_5\times S^5$ space-time in the presence of a constant $B$-field, which was used in ref.~\cite{Filev:2007gb} to study magnetic catalysis of chiral symmetry breaking:
\begin{eqnarray}
ds^2 &=-& \frac{u^2}{R^2}(-dt^2 + dx_1^2 + dx_2^2 + dx_3^2) + \frac{R^2}{u^2}du^2 + R^2(d\theta^2 + \cos^2\theta d\Omega_3^2 + \sin^2\theta d\phi^2)\ , \nonumber \\
B_{(2)} &=& H dx_1\wedge dx_2\ , ~~~ e^{\Phi} = g_s
\end{eqnarray}
It is convenient to change coordinates to:
\begin{equation}
\rho = u\cos\theta, ~~~L = u\sin\theta\ 
\end{equation}
and consider an ansatz where the D7--brane wraps $t, x_1, x_2, x_3, \rho, \Omega_3$ directions and has a nontrivial profile $L=L(\rho)$. Substituting into the action (\ref{DBI}) and Wick rotating, we get:
\begin{equation}
S_{DBI}^{E} = \frac{2\pi^2\mu_7}{g_s}\beta V_{3}\int\limits_0^\infty d\rho \rho^3\sqrt{1+\frac{R^4\,H^2}{(\rho^2+L(\rho)^2)^2}}\,\sqrt{1+L'(\rho)^2}\ ,
\end{equation}
where $\beta$ is the period of the Euclidean time $\tau$, while $V_3$ comes from integrating along the $x_1, x_2, x_3$ directions (one can think of these directions as spanning a torus if finite $V_3$ is needed).  The free energy density can then be written as:
\begin{equation}
{\cal F} = \frac{2\pi^2\mu_7}{g_s} I_{D7} ,
\end{equation}
where:
\begin{equation}\label{I7}
I_{D7} = \int\limits_0^{\rho_{max}} d\rho\left\{ \rho^3\sqrt{1+\frac{R^4\,H^2}{(\rho^2+L(\rho)^2)^2}}\,\sqrt{1+L'(\rho)^2} -\rho\sqrt{1+\rho^4}\right\} .
\end{equation}
Note that the second term in (\ref{I7}) was obtained by subtracting the free energy of the trivial $L(\rho)\equiv 0$ embedding. We have also introduced a UV cutoff $\rho_{max}$.

Holography dictates that the asymptotic of the solution $L(\rho)$ at large $\rho$ should be:
\begin{equation}\label{Linf}
L(\rho) = m + \frac{c}{\rho^2} +\dots
\end{equation}
and one can check that integral in (\ref{I7}) is finite. Furthermore, the constants $m$ and $c$ are proportional to the bare mass and fundamental condensate in the dual theory. \footnote{The exact relation \cite{Filev:2007gb} to the bare mass $m_q$ is $m = \frac{(2\pi\alpha')\,m_q}{\sqrt{B}}$, where $B$ is the external magnetic field. }. In ref.~\cite{Filev:2007gb} it was shown that this set-up features dynamical mass generation and chiral symmetry breaking induced by the external magnetic field.  

\begin{figure}[t]
\begin{center}
\begin{tikzpicture}[x=4.8cm, y=2.4cm, every node/.style={scale=1.1}]
  \node[circle,draw,minimum size=1cm] (I1) at (0,0) {$\rho$};

  \foreach \i in {1,...,5} {
    \node[circle,draw,fill=blue!10,minimum size=0.7cm] (H1\i) at (1,1.8-\i*0.9) {};
  }
  \node[above=0.2cm of H13] {\scriptsize Linear + Tanh};

  \foreach \i in {1,...,5} {
    \node[circle,draw,fill=green!10,minimum size=0.7cm] (H2\i) at (2,1.8-\i*0.9) {};
  }
  \node[above=0.2cm of H23] {\scriptsize Linear + Tanh};

  \node[circle,draw,minimum size=1cm,fill=red!10] (O1) at (3,0) {$g(\rho)$};
  \node[below=0.3cm of O1] {\scriptsize Linear};

  \foreach \i in {1,...,5} {
    \draw[->,thick] (I1) -- (H1\i);
  }

  \foreach \i in {1,...,5} {
    \foreach \j in {1,...,5} {
      \draw[->,opacity=0.3] (H1\i) -- (H2\j);
    }
  }

  \foreach \i in {1,...,5} {
    \draw[->,thick] (H2\i) -- (O1);
  }
\end{tikzpicture}
\end{center}
\caption{Architecture of the ANN used to learn the profile $L(\rho)$. The parameter $\rho$ is in the interval $[0, \rho_{max}]$ and $L(\rho) = m + (\rho_{max} - \rho) g(\rho)$ }
\label{fig:ANN1}
\end{figure}
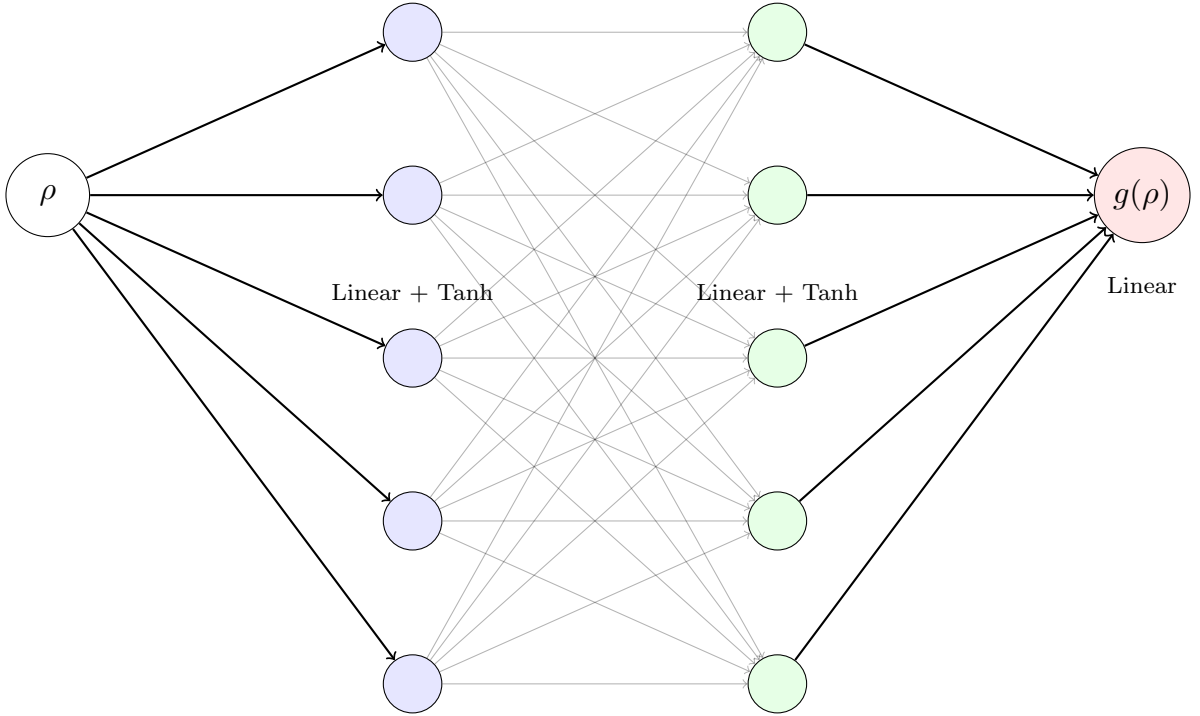

Our next goal is to construct an artificial neural network (ANN) to learn the profile of the embedding $L(\rho)$ by minimizing the free energy functional (\ref{I7}). In figure~\ref{fig:ANN1} we have presented a schematic representation of the ANN architecture used to learn the profile.  

The figure shows a feedforward artificial neural network (ANN) with a single input $\rho$ and a scalar output $g(\rho)$. The network consists of two hidden layers. Although the diagram displays only five neurons per hidden layer for visual simplicity, the actual network used in practice has at least ten neurons in each hidden layer. Each hidden layer applies a linear transformation followed by a hyperbolic tangent (tanh) activation function. The final output layer performs a linear transformation without a nonlinearity to produce the output $g(\rho)$. All neurons are fully connected between layers, making the architecture well-suited for approximating smooth scalar functions of a single input variable. 

To impose the boundary condition that $L(\infty) = m$ (see equation (\ref{Linf})), we select a UV cutoff $\rho_{max}$ and we perform one final linear transformation, relating the profile function $L(\rho)$ to the output of the ANN $g(\rho)$:
\begin{equation}
L(\rho) = m + (\rho_{max} - \rho)g(\rho)
\end{equation}

To implement this architecture in practice, we use PyTorch, a modern automatic differentiation and neural network library. \footnote{The codebase is available at \url{https://github.com/vesofilev/pinn_d7}.}
The network $g(\rho)$ is represented by a torch.nn sequential model with two hidden layers, each containing at least 10 neurons and using the hyperbolic tangent (tanh) activation function. The output layer is linear, and the full ansatz for $ L(\rho)$ is enforced explicitly as $L(\rho) = m + (\rho_{\text{max}} - \rho)g(\rho)$, guaranteeing the UV boundary condition $L(\rho_{\text{max}}) = m$ is satisfied by construction.

The free energy functional (\eqref{I7}) is then discretized over a grid in $\rho$, and the derivative $L’(\rho)$ is computed using PyTorch’s automatic differentiation engine. This allows us to treat the regularized DBI action as a differentiable loss function, which is minimized using the Adam optimizer. The neural network parameters are updated through gradient descent to extremize the action and find the optimal brane embedding. To improve numerical efficiency, we use a nonuniform grid for $\rho$, with denser sampling near the IR region where the profile varies rapidly. This approach avoids solving the Euler–Lagrange equations directly and instead recasts the variational problem as a machine learning task: the network “learns” the brane profile by minimizing the physical free energy.

\begin{figure}[t]
\begin{center}
\includegraphics[scale=1.1]{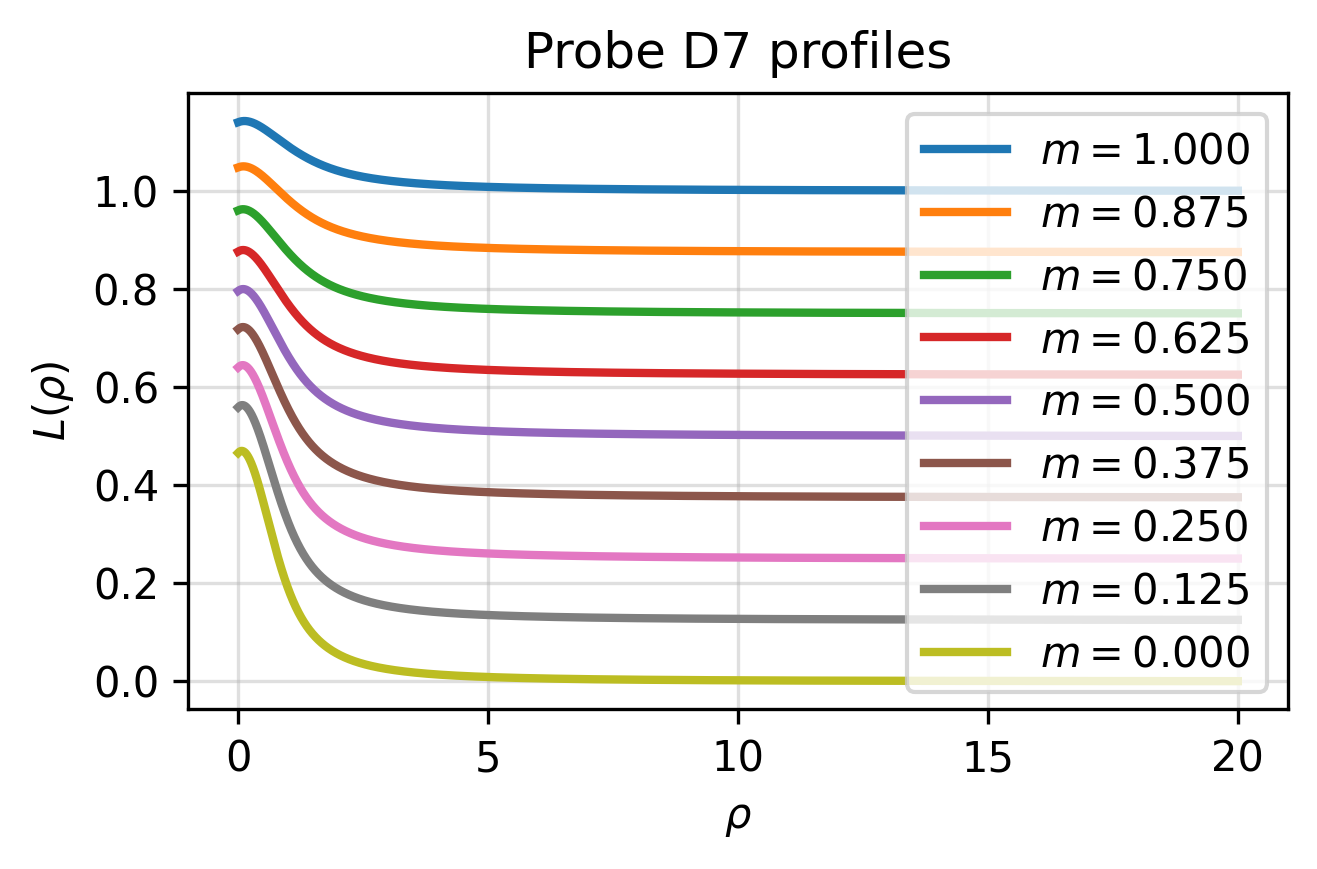}
\caption{Plots of $L$ versus $\rho$ for various mass parameter $m$. One can see that the ANN produces smooth functions.}
\label{fig:embeddings}
\end{center}
\end{figure}

\begin{figure}[h]
\begin{center}
\includegraphics[scale=0.8]{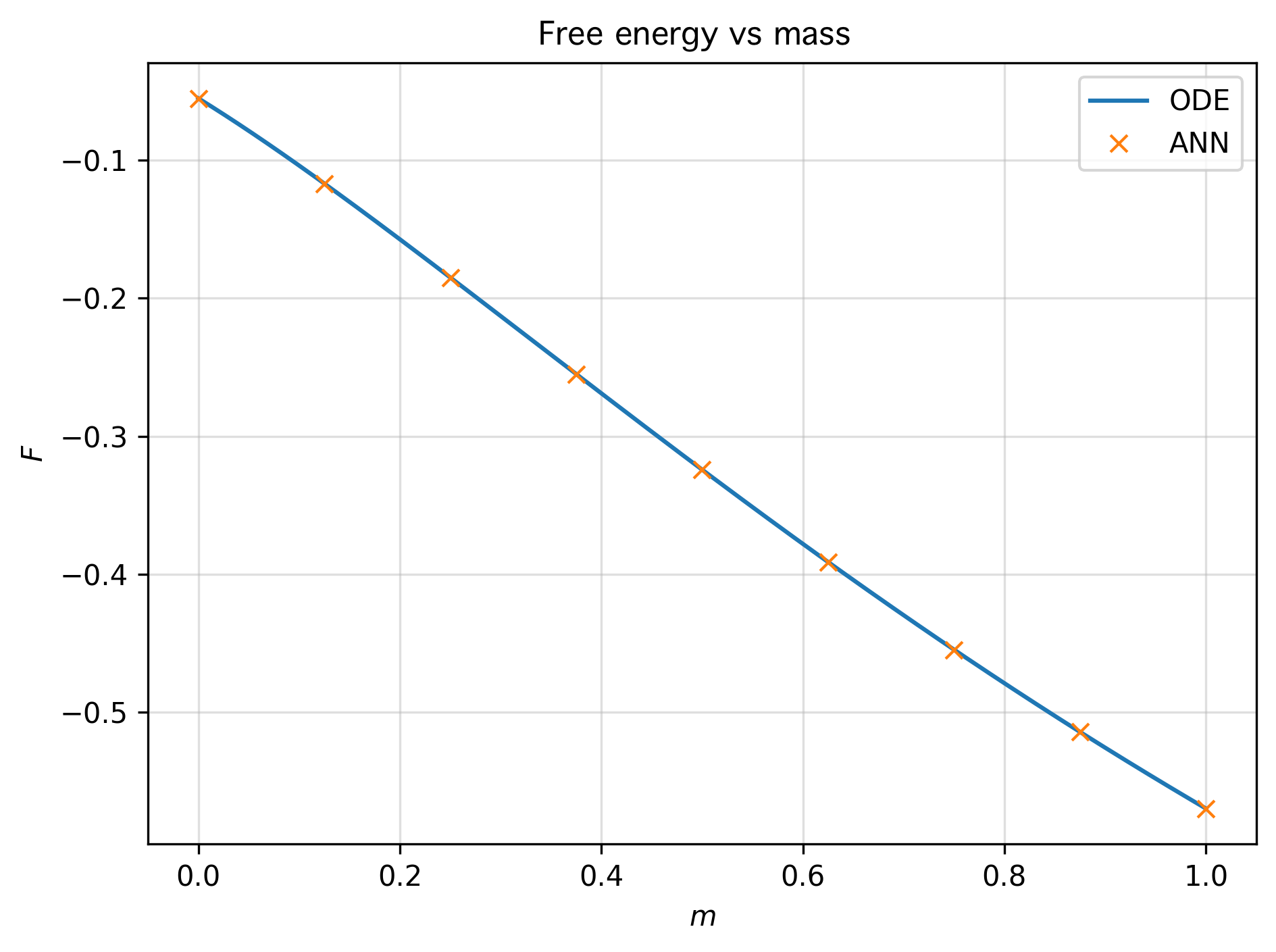}
\caption{Comparison of the free energy functional $I_{D7}$ (denoted by $F$ in the plot) between the numerical solution obtained by solving the equations of motion using shooting techniques and the free energy as given by the loss of the training process.  One can observe an excellent agreement.}
\label{fig:F}
\end{center}
\end{figure}

In Figures \ref{fig:embeddings} and \ref{fig:F}, we present the embedding profiles for different values of the mass parameter $m$, as well as a comparison of the free energy functional $I_{D7}$ obtained both by numerically solving the ODEs and by training the ANN (with the architecture shown in Figure \ref{fig:ANN1}).

We restrict the dimensionless parameter to $m \in [0,1]$ primarily for convenience. The physically interesting effect of the magnetic field occurs at $m=0$, where dynamical chiral symmetry breaking can be observed. For $m>1$, no additional phase transitions occur. Moreover, for large $m$, accurate approximate solutions (expansions in $1/m$) can be obtained. Finally, keeping $m \sim 1$ ensures that we remain in the regime $m \ll \rho_{\max}$, well below the UV cutoff, where the approximation $L(\rho_{\max}) \approx m$ remains valid.

As one can see the ANN produces smooth embeddings with free energies matching those obtained by solving numerically the EOM using shooting techniques.  The plots have been produced by using 60000 epochs to train a separate neural network for each mass and the relative difference between the free energies is of order $10^{-5}$.  

We rely on PyTorch’s default initialisation for fully connected layers, which gives moderately scaled weights and small, near‑zero biases. Because the UV boundary condition is hard‑wired into the parameterisation, the network could even start with all weights set to zero and still satisfy the constraint exactly. In practice, we include small random biases to break symmetry and avoid flat starts, which improves early gradient flow and stabilises training without steering the model toward any specific embedding shape.

Note that training a separate network for each mass is not an optimal strategy since one would expect that the weights of the neural network would change smoothly as the mass parameter is changed. Indeed, in section~\ref{LandM} we will exploit this intuition to train a network that learns the whole one parameter family of embeddings $L(\rho,m)$ but before that we will look into one more example of a holographic set-up: that of the meson melting phase transition.

\subsection{Learning in meson melting}
In this section we provide an example of learning the embeddings for the holographic meson melting set-up in the D3/D7 brane system first studied in ref.~\cite{Babington:2003vm} (see also refs.~\cite{Mateos:2006nu} and \cite{Albash:2006ew}).  We will briefly review the set-up, we start with the AdS-Schwarzschild background:
\begin{equation}\label{BH-u}
ds^2 = -\frac{u^4-u_0^4}{u^2R^2}dt^2 + \frac{u^2}{R^2}(dx_1^2+dx_2^2+dx_3^2) + \frac{u^2R^2}{u^4-u_0^4}du^2+ R^2(d\theta^2 + \cos^2\theta d\Omega_3^2 + \sin^2\theta d\phi^2)\ .
\end{equation}
It is convenient \cite{Babington:2003vm} to introduce a coordinate transformation which restores the conformally flat $\mathbb{R}^6$ part of the geometry. We apply first:
\begin{equation}
u = \frac{\sqrt{r^4 + r_0^4}}{r}, ~~~u_0 = \sqrt{2}\,r_0\ .
\end{equation}
and then define the radial coordinates in $\mathbb{R}^4$ and $\mathbb{R}^2$ respectively:
\begin{equation}
\rho = r\,\cos\theta, ~~~L = r\,\sin\theta\ .
\end{equation}
Considering again a D7--brane embedding wrapping the $t, x_1, x_2, x_3, \rho, \Omega_3$ directions, for the Wick rotated DBI action we obtain:
\begin{equation}
S_{DBI}^{E} = \frac{2\pi^2\mu_7}{g_s}\beta V_{3}\int\limits_{\rho_{min}}^\infty d\rho \rho^3\left(1-\frac{r_0^8}{(\rho^2+L(\rho)^2)^4}\right)\,\sqrt{1+L'(\rho)^2}\ ,
\end{equation}
while for the free energy functional $I_{D7}$ we get:
\begin{equation}\label{I7hot}
I_{D7} = \int\limits_{\rho_{min}}^{\rho_{max}} d\rho\left\{ \rho^3\left(1-\frac{r_0^8}{(\rho^2+L(\rho)^2)^4}\right)\,\sqrt{1+L'(\rho)^2} -\rho^3\right\} + \frac{\rho_{min}^4}{4} ,
\end{equation}
where we have again regulated by subtracting a boundary term $\rho_{max}^4/4$. Note that in this case we don't have logarithmic divergences. Depending on the topology of the embedding (either reaching the black hole horizon or closing above it) the value of $\rho_{min}$ is either zero for embeddings closing above the horizon (Minkowski embedding) or $\rho_{min}$ is determined from:
\begin{equation}
\rho_{min}^2 + L(\rho_{min})^2 = r_0^2\ ,
\end{equation}
for black hole embeddings.

It is well established \cite{Babington:2003vm} that the topology change transition between the two classes of embeddings (Minkowski and black hole) correspond to a meson melting phase transition in the dual field theory where the Minkowski embeddings are associated with the confined phase and black hole embedding with the deconfined phase.  It was further established in refs. ~\cite{Mateos:2006nu, Albash:2006ew} that the meson melting phase transition is of first order. This leads to multivaluedness of the free energy versus mass dependence near the phase transition and requires further care when training the neural network. 

To avoid the complication of determining $\rho_{min}$ for the case of black hole embeddings we rewrite the free energy functional (\ref{I7hot}) as:
\begin{equation}\label{ID7-hot}
I_{D7} = \int\limits_{0}^{\rho_{max}} d\rho\left\{ \rho^3\,{\rm max}\left[\left(1-\frac{r_0^8}{(\rho^2+L(\rho)^2)^4}\right), 0\right]\,\sqrt{1+L'(\rho)^2} -\rho^3\right\}\ ,
\end{equation}
which avoids the explicit specification of $\rho_{min}$ and we always integrate from $\rho_{min}=0$ but clamp the contribution to the free energy from the area below the horizon to zero. Note also that the last term in (\ref{I7hot}) is also taken care of. This trick allows us to use exactly the same training neural network architecture and training strategy as in the previous section. We don't specify the topology of the embedding and let the algorithm decide which kind of topology is energetically more favorable. As mentioned above this becomes sensitive near the first order phase transition where the metastable phase occurs as a local minimum for the loss function, but training a ANN for each mass (which we do in this section) and increasing the number of training epochs near the phase transition allows us to properly map the first order phase transition pattern.

\begin{figure}[h]
\begin{center}
\includegraphics[scale=0.9]{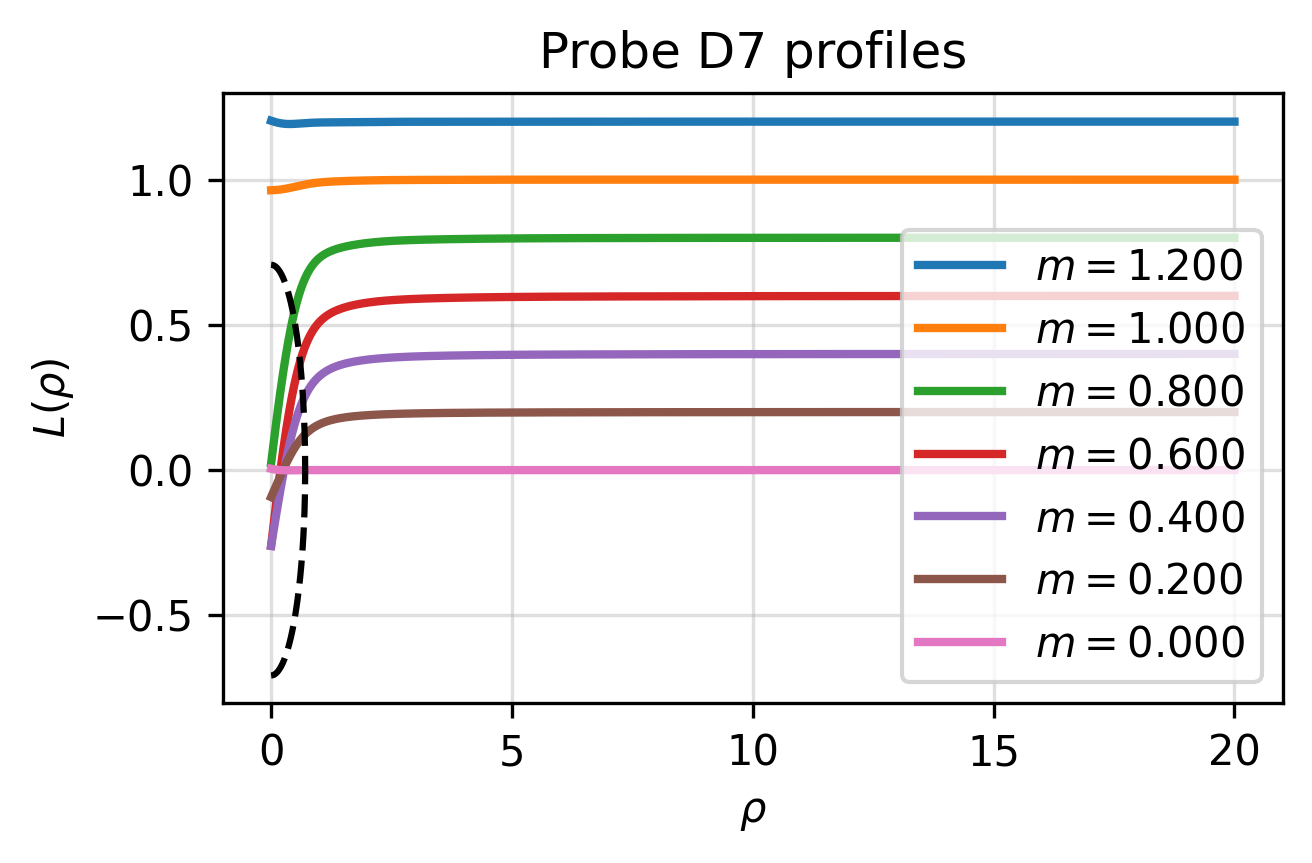}
\caption{Plots of $L$ versus $\rho$ for various mass parameter $m$. The curves inside the horizon are not physical and do not contribute to the gradient during training. The red rectangle represent the area of the first order phase transition pattern.}
\label{fig:embeddings-hot}
\end{center}
\end{figure}

In figure~\ref{fig:embeddings-hot}  we have presented plots of D7--brane embeddings for a range of different bare mass parameters $m$. The part of the embeddings below the horizon are not physical and do not contribute to the gradient during training. One can clearly observe the presence of the two classes of embeddings: Minkowski and Black Hole embedding. 

\begin{figure}[h]
\begin{center}
\includegraphics[scale=0.9]{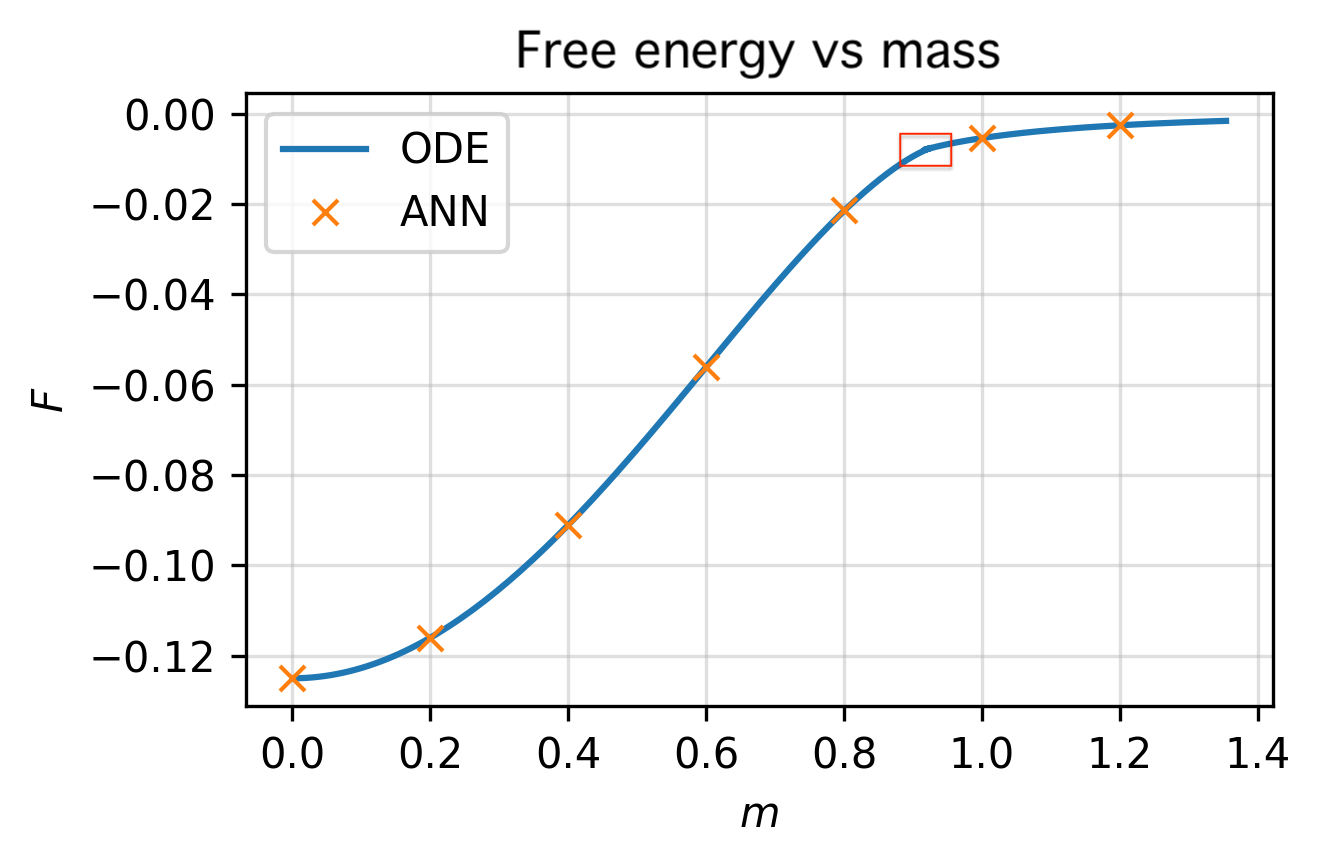}
\caption{Plot of the free energy functional  (\ref{ID7-hot}) as a function of the bare mass parameter $m$. The agreement between the two approaches is excellent.}
\label{fig:F-hot}
\end{center}
\end{figure}

In figure~\ref{fig:F-hot} we have presented a plot of the free energy functional (\ref{ID7-hot}) as function of the bare mass parameter $m$ using both numerical ODE solver (Mathematica) and our ANN approximation.  One can observe an excellent agreement between the two approaches.  The red rectangle in the figure represents the area of the first order phase transition.

\begin{figure}[h]
\begin{center}
\includegraphics[scale=0.9]{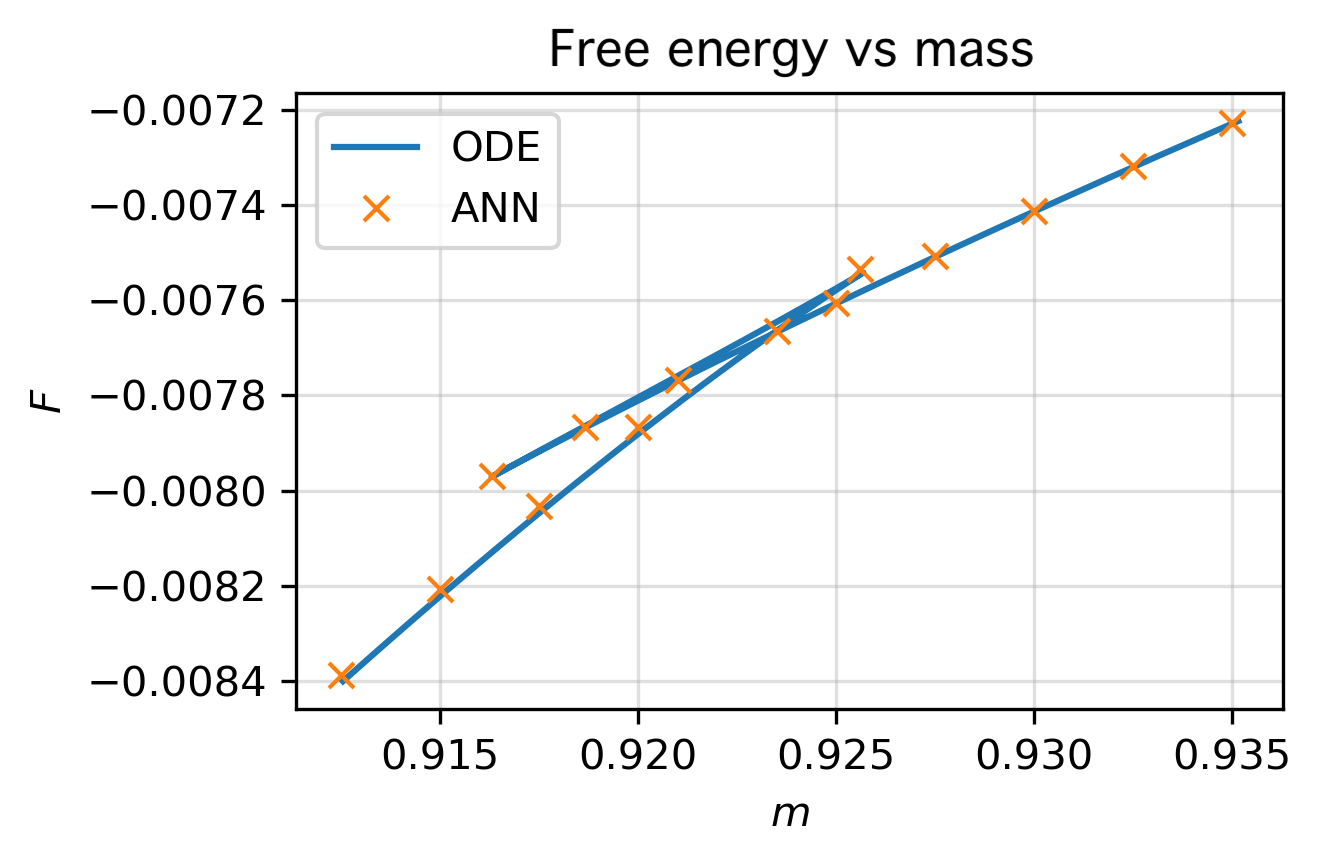}
\caption{Plot of the free energy functional  (\ref{ID7-hot}) as a function of the bare mass parameter $m$ near the phase transition at $m \approx 0.9234$, corresponding to the red rectangle in figure~\ref{fig:F-hot}. The agreement between the two approaches if excellent.}
\label{fig:F-hot-zoom}
\end{center}
\end{figure}

To test the accuracy of our ANN technique in figure~\ref{fig:F-hot-zoom} we generate the comparison between the two approaches near the critical parameter $m_{cr}\approx 0.9234$. As one can see the accuracy of the ANN method is sufficient to resolve the pattern near the phase transition. One can see that the agreement is better in the Minkowski phase, one reason for this is the smaller multivalued region. In fact, if when training the ANNs for different masses one starts with an initial state for the ANN given by one of the previous masses then the ANN traces also the metastable phases. That is in the multivalued region the optimizer cannot change the topology class of the embedding. To generate the plots in figure~\ref{fig:F-hot-zoom} we have started with embeddings in Black Hole phase for small masses and slowly increased the mass as approaching the critical mass. Once we reached the maximum possible mass in the black hole phase (metastable) we considered a relatively larger mass ($m \approx 0.935$) corresponding to a Minkowski embedding which is far from the multivalued region and then slowly decreased the masses passing through the critical mass and going down all the way to the metastable Minkowski phase. As one can see the proposed ANN method is useful for branch tracking.

Note that the ODE-generated curve includes also the unstable free-energy solution in the multivalued region, which the ANN does not reproduce. For any generic point in the multivalued region (i.e., excluding turning points and the critical point) there are three possible free energies: the lowest corresponds to the stable phase (which the ANN can capture); the middle corresponds to a metastable phase (which the ANN can capture); and the highest corresponds to an unstable phase (which the ANN cannot capture). The reason the ANN method cannot capture the unstable phase is that, in the multivalued region, the three stationary solutions correspond to two minima (global/stable and local/metastable) and one local maximum (unstable) of the regularized DBI functional; gradient-based minimization naturally converges to minima and not to the unstable maximum.

\section{Learning a one parameter family of embeddings}\label{LandM}

In the previous section we trained a different neural network for every bare mass parameter $m$ that we've considered. This suggests that if we want to generate a dense curve of the free energy versus the bare mass we have to dedicate considerably more computational  resources than if we solve numerically the equations of motion. However,  if we learn the whole one parameter family of embeddings $L(\rho, m)$ we can obtain a smooth approximation of the for every value of $m$ and even take derivatives with respect to $m$.  Furthermore, the singe-mass strategy from the previous section is sub-optimal from numerical point of view: for every value of the bare mass parameter $m$ we initialize a fresh network and fully train it, we can of course use as a starting point the network for previous mass which was close enough but we still create a full copy of the network while just a small modification would suffice. Therefore, it is natural to consider a network where the weights would change continuously with the bare mass parameter.

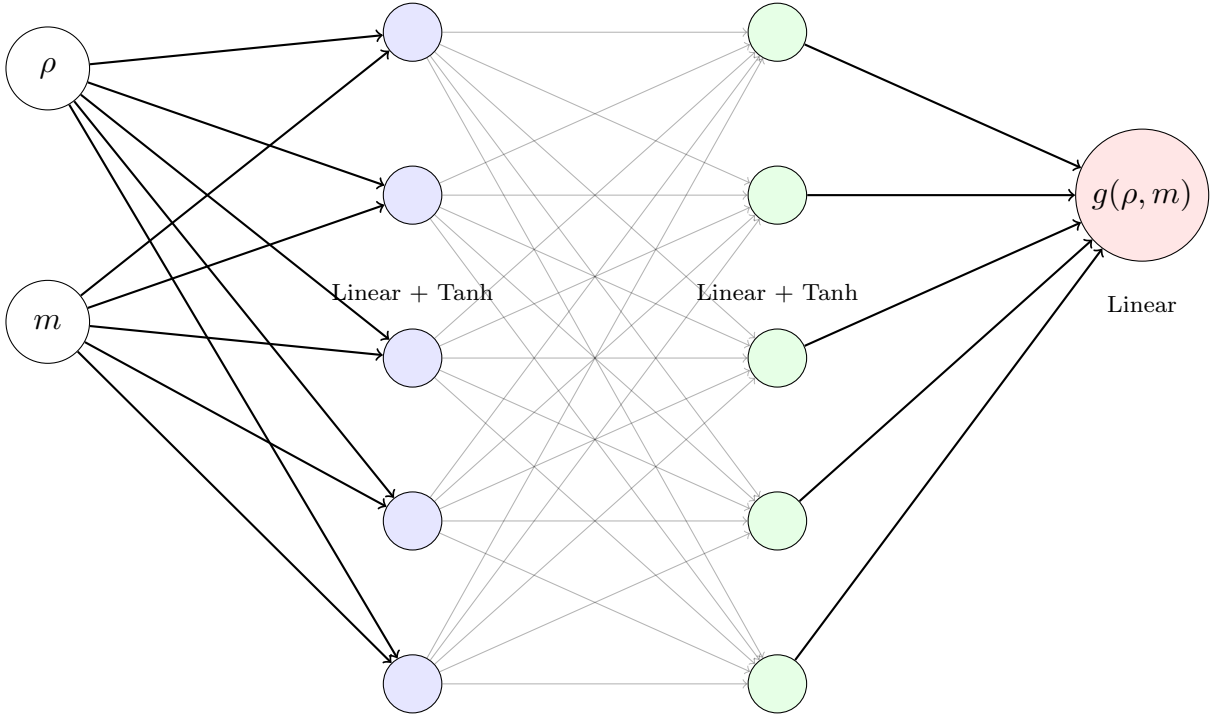
\begin{figure}
\begin{center}
\begin{tikzpicture}[x=4.8cm, y=2.4cm, every node/.style={scale=1.1}]
  \node[circle,draw,minimum size=1cm] (I1) at (0,0.7) {$\rho$};
  \node[circle,draw,minimum size=1cm] (I2) at (0,-0.7) {$m$};

  \foreach \i in {1,...,5} {
    \node[circle,draw,fill=blue!10,minimum size=0.7cm] (H1\i) at (1,1.8-\i*0.9) {};
  }
  \node[above=0.2cm of H13] {\scriptsize Linear + Tanh};

  \foreach \i in {1,...,5} {
    \node[circle,draw,fill=green!10,minimum size=0.7cm] (H2\i) at (2,1.8-\i*0.9) {};
  }
  \node[above=0.2cm of H23] {\scriptsize Linear + Tanh};

  \node[circle,draw,minimum size=1cm,fill=red!10] (O1) at (3,0) {$g(\rho, m)$};
  \node[below=0.3cm of O1] {\scriptsize Linear};

  \foreach \i in {1,...,5} {
    \draw[->,thick] (I1) -- (H1\i);
    \draw[->,thick] (I2) -- (H1\i);
  }

  \foreach \i in {1,...,5} {
    \foreach \j in {1,...,5} {
      \draw[->,opacity=0.3] (H1\i) -- (H2\j);
    }
  }

  \foreach \i in {1,...,5} {
    \draw[->,thick] (H2\i) -- (O1);
  }
\end{tikzpicture}
\end{center}
\caption{Architecture of the ANN used to learn the one parameter family $L(\rho, m)$. The parameter $\rho$ is in the interval $[0, \rho_{max}]$ and $L(\rho, m) = m + (\rho_{max} - \rho) g(\rho, m)$ }
\label{fig:ANN2}
\end{figure}

To this end we consider the architecture in figure~\ref{fig:ANN2}. The figure shows a feedforward artificial neural network (ANN) with a two parameter input $\rho, m$ and a scalar output $g(\rho, m)$.  The fact that we use two input parameters as opposed to one is the only architectural difference with the neural network from figure~\ref{fig:ANN1}. The network again consists of two hidden layers. Each hidden layer representing a linear transformation followed by a hyperbolic tangent (tanh) activation function. The final output layer performs a linear transformation without a nonlinearity to produce the output $g(\rho, m)$. All neurons are fully connected between layers, making the architecture well-suited for approximating smooth scalar functions of two variables. 
To impose the boundary condition that $L(\infty) = m$ (see equation (\ref{Linf})), we select a UV cutoff $\rho_{max}$ and we perform one final linear transformation, relating the profile function $L(\rho, m)$ to the output of the ANN $g(\rho, m)$:
\begin{equation}
L(\rho, m) = m + (\rho_{max} - \rho)g(\rho, m)
\end{equation}

We use a deliberately small, zero‑mean Gaussian initialization for the weights so the model begins as a gentle perturbation around the bare mass input. As with the single‑mass case, the architectural parameterization ensures that the boundary condition is satisfied even if all weights were zero, so the network starts from a valid physical state. Small random weight perturbations (with biases set to zero) provide slight asymmetry and enhance numerical stability and convergence, while preserving the intended inductive bias toward smooth dependence on the mass parameter.

The conditional training routine first builds a non-uniform collocation grid
\begin{equation}
\rho_0=0<\rho_1<\dots<\rho_{N-1}=\rho_{\max},
\end{equation}
obtained by concatenating a dense sub-grid on $[0,\rho_{\rm cut}]$ with a coarser one on $[\rho_{\rm cut},\rho_{\max}]$.
This concentrates collocation points in the IR region where the embeddings vary most rapidly while keeping the total number of points moderate (default: $N=2000$).

At every optimization step we sample a \emph{mini–batch} of $N_m$ distinct masses,
\begin{equation}
m^{(1)},\dots,m^{(N_m)}\stackrel{\text{i.i.d.}}{\sim}\mathcal{U}(m_{\min},m_{\max}),
\end{equation}
and evaluate the regularised DBI functional $\mathcal{F}[L(,\cdot,,m^{(i)})]$ on the whole $\rho$–grid using automatic differentiation to obtain~$\partial_\rho L$.
The loss is the average free energy over the $N_m$~masses:
\begin{equation}
\label{eq:cond_loss}
\mathcal{L}(\theta)=\frac1{N_m}\sum_{i=1}^{N_m} \mathcal{F}\bigl[L_{\theta}(m^{(1)},\dots,m^{(N_m)})\bigr] .
\end{equation}
Note that because of the random sampling there are no cancellations of gradients in the sum and the minimum of the average is achieved at the minimum for a given mass.   While increasing the batch size allows the network to adapt to more masses in a single step, the evaluation is also somewhat slower. We found that even with a sample size of $N_m = 1$ (that is we sample one mass a time) we have sufficiently fast convergence while keeping the evaluation fast enough. Furthermore, to accelerate convergence and make sure that the network learns the edge of the mass interval we additionally inject, with some fixed probability (usually $20\%$) an \emph{extremal} mass $m_{\min}$ or $m_{\max}$ in the batch.

\vspace{2pt}
The parameters $\theta$ are optimised with Adam ($\mathrm{lr}=5\times10^{-5}$ by default) usually with $6\times 10^{4}-1.2\times 10^5$ epochs. Every $10^{3}$ steps the training loop (i) saves a checkpoint (network weights and optimiser state) and (ii) evaluates $F(m)$ on a dense evaluation grid of masses, which can be monitored in real time.

\begin{figure}
\begin{center}
\includegraphics[scale=1.0]{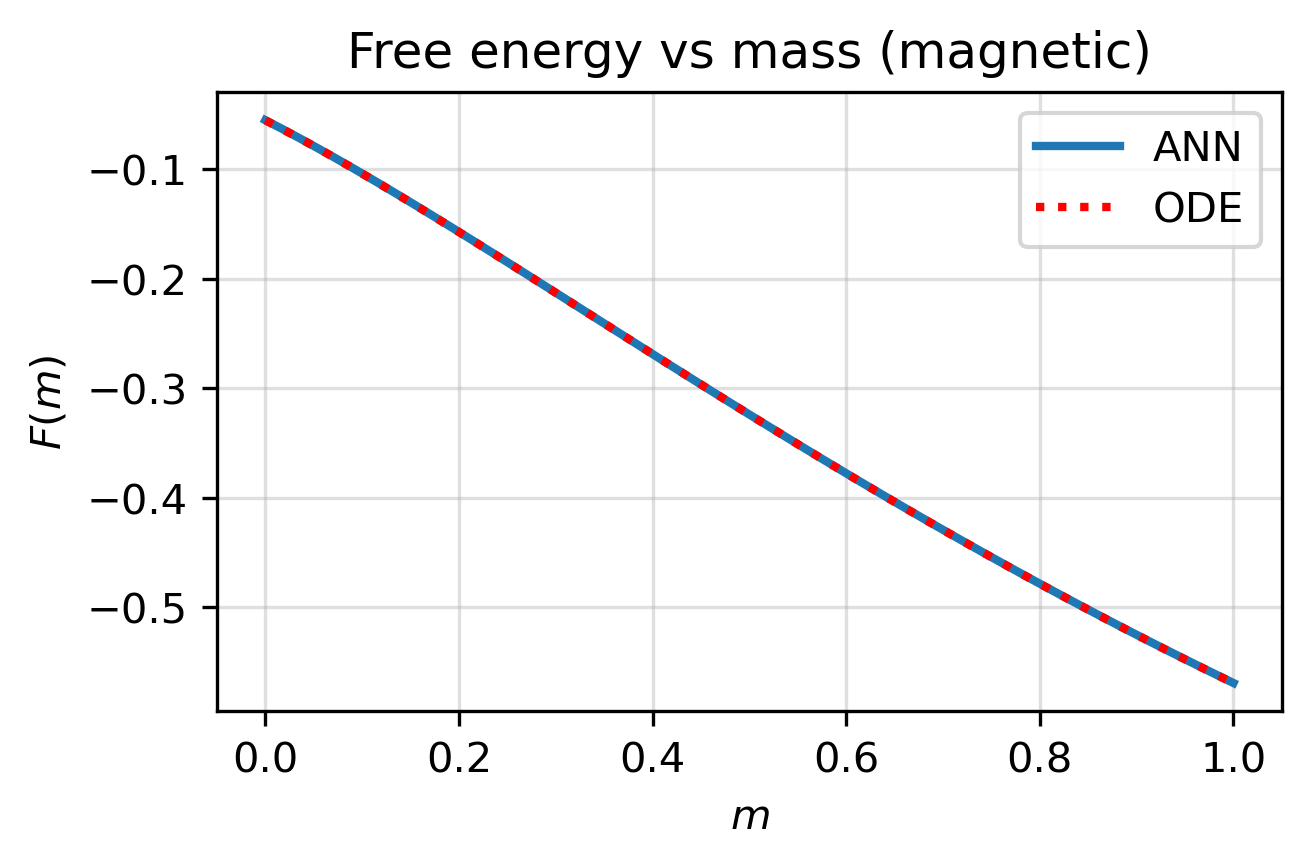}
\end{center}
\caption{Comparison of $F(m)$ between the ODE and ANN methods for the functional (\ref{I7}).  The ANN curve was obtained with 120000 epochs.}
\label{fig:F-mag-condit}
\end{figure}

\begin{figure}
\begin{center}
\includegraphics[scale=1.0]{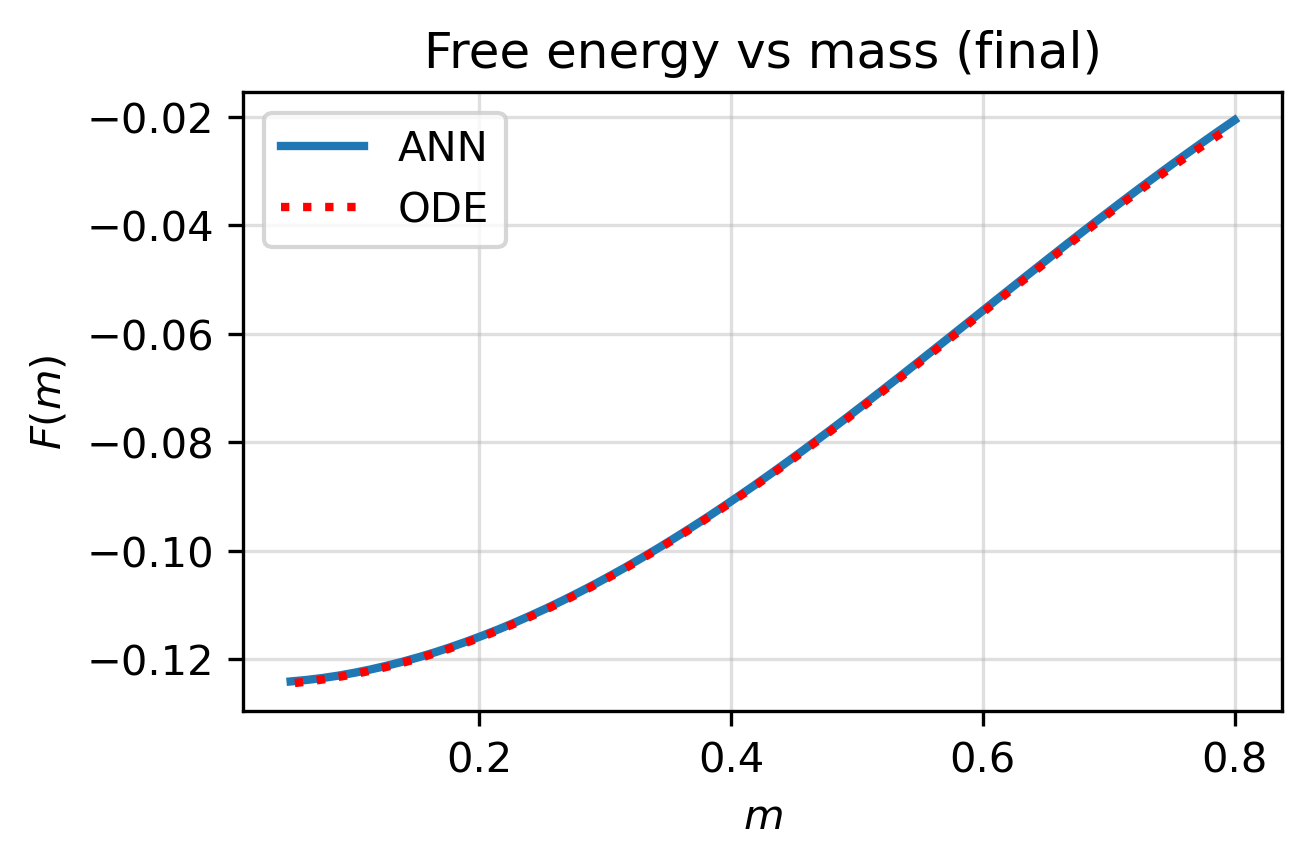}
\end{center}
\caption{Comparison of $F(m)$ between the ODE and ANN methods for the functional (\ref{I7hot}).  The ANN curve was obtained with 80000 epochs.}
\label{fig:F-hot-condit}
\end{figure}

In figure~\ref{fig:F-mag-condit} and figure~\ref{fig:F-hot-condit} we have presented plots of the free energy versus bare mass curve for both the finite external magnetic field and finite temperature setups considered in the previous section. One can observe the perfect agreement between the ODE and ANN results.  In the finite temperature case training the network near the phase transition becomes challenging because of the presence of numerically close metastable phases, this is why we train the conditional network only in a given phase (the deconfined phase was presented in the plot).

\begin{figure}
\begin{center}
\includegraphics[scale=1.0]{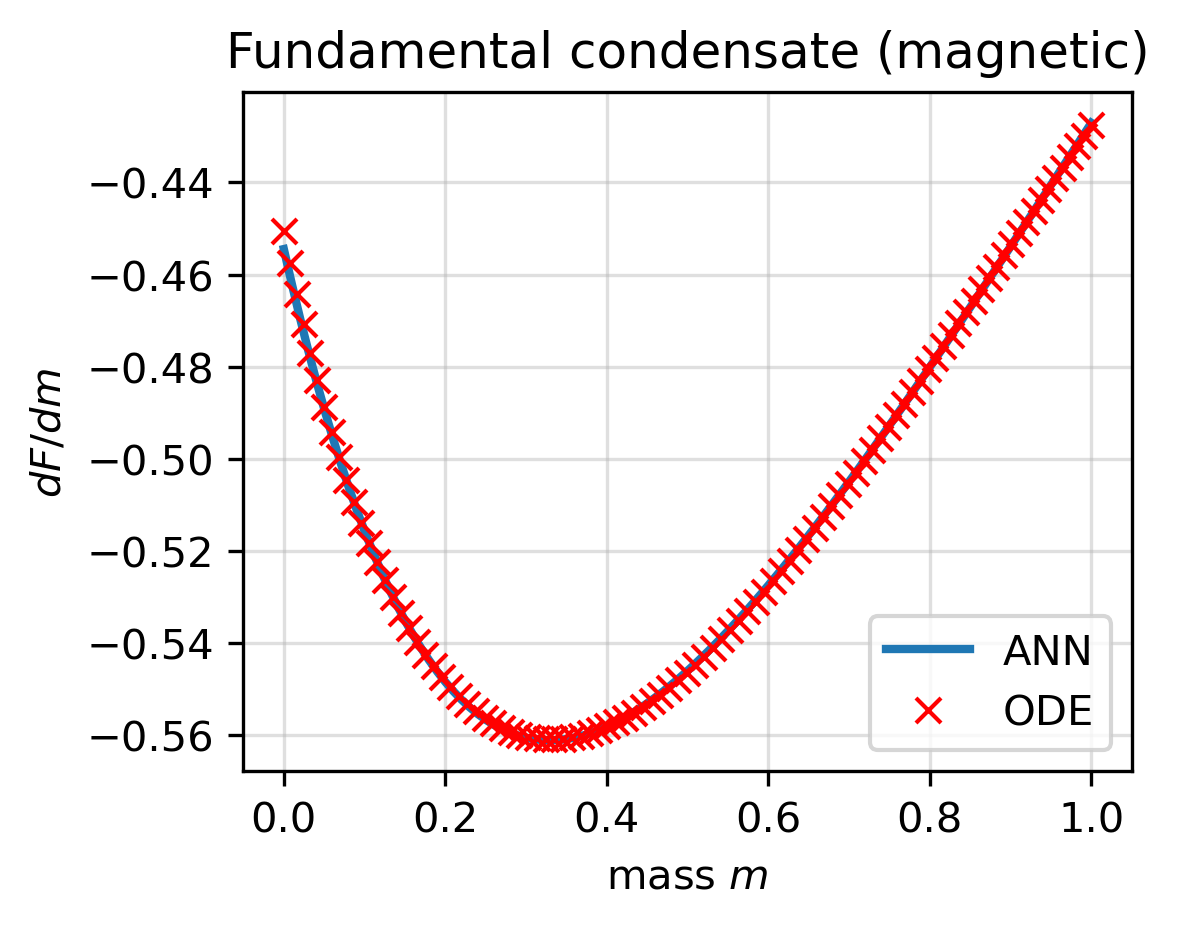}
\end{center}
\caption{Comparison of condensate curve $\partial_mF(m)$ between the ODE and ANN methods for the functional (\ref{I7}).  The ANN curve was obtained with 120000 epochs. Even though a derivative is taken, the agreement is still excellent.}
\label{fig:c-vs-m}
\end{figure}

Because $m$ is an input variable, the trained network provides direct access to the derivative
\begin{equation}
c(m)=\frac{\partial\mathcal{F}}{\partial m}
=
\frac{\partial}{\partial m}
\int_0^{\rho_{\max}}{\cal L}_{\rm DBI}\bigl[\rho,L(\rho,m)\bigr]d\rho ,
\end{equation}
which is proportional to the fundamental condensate.
In practice we compute $\partial_m\mathcal{F}$ by a single call to
\texttt{torch.autograd.grad}  thus obtaining the whole curve $c=c(m)$ at a negligible additional cost. In figure~\ref{fig:c-vs-m} we have presented the condensate versus bare mass curve for the functional (\ref{I7}). One can see that despite the derivative the agreement between the ODE and ANN approached remains excellent.

\section{Inverse Problem: Learning the Geometry}
In this section we will solve an ``inverse problem'' in the spirit of ref.~\cite{Hashimoto:2018ftp} where the authors considered deep learning techniques to learn certain aspects of the geometry (parts of the metric) given data from the dual theory.  

We start with the ansatz for the metric:
\begin{eqnarray}
  ds^2 &=& - f_1(r) dt^2 + f_2(r)(dx_1^2 + dx_2^2 + dx_3^2) + \frac{R^2}{r^2} dr^2 + R^2 d\Omega_5^2\ ,  \\
  d\Omega_5^2 &=& d\theta^2 + \cos^2\theta\, d\Omega_3^2 + \sin^2\theta\, d\phi^2 \nonumber
\end{eqnarray}
where $f_1(r)$ and $f_2(r)$ are undetermined functions with asymptotic behavior:
\begin{equation}\label{asymp_f12}
f_1(r) = \frac{r^2}{R^2} + \dots \ ,\, ~~~f_2(r) = \frac{r^2}{R^2} + \dots\ .
\end{equation}
ensuring that the geometry is asymptotically AdS$_5\times S^5$. Note that we have used the freedom of selecting a radial variable to fix the $g_{rr}$ component of the metric. With this choice of radial variable we can again define: 
\begin{equation}
\rho = r\,\cos\theta \ , ~~~L=r\,\sin\theta\ .
\end{equation}
and select a D7--brane embedding wrapping the time $t$ and space $x_1, x_2, x_3$ directions as well as the radial $r$ and $\Omega_3$ directions. For the DBI action we obtain:
\begin{equation}\label{I7_V}
  \int\limits_{\rho_0}^{\infty} d\rho\; \rho^3 \left[V\left(\sqrt{\rho^2 + L(\rho)^2}\right) \sqrt{1 + L'(\rho)^2} - 1\right]\ ,
\end{equation}
where the potential is given by:
\begin{equation}
V(r) = \frac{\sqrt{f_1(r)} \cdot f_2(r)^{3/2}}{r^4}\ .
\end{equation}
Note that the asymptotic behavior of $f_1$ and $f_2$ in equation (\ref{asymp_f12}) suggests that we have:
\begin{equation}
V(r) = 1 + \dots \ ,
\end{equation}
this together with the asymptotic behavior of $L(\rho)$:
\begin{equation}
L(\rho) = m + \frac{c}{\rho^2} + \dots .
\end{equation}
ensures that the last term in the brackets in (\ref{I7_V}) cancels all of the divergences and the integral is convergent. 

\begin{figure}[t]
\begin{center}
\begin{tikzpicture}[x=4.8cm, y=2.4cm, every node/.style={scale=1.1}]
  \node[circle,draw,minimum size=1cm] (I1) at (0,0) {$r$};

  \foreach \i in {1,...,5} {
    \node[circle,draw,fill=blue!10,minimum size=0.7cm] (H1\i) at (1,1.8-\i*0.9) {};
  }
  \node[above=0.2cm of H13] {\scriptsize Linear + Tanh};

  \foreach \i in {1,...,5} {
    \node[circle,draw,fill=green!10,minimum size=0.7cm] (H2\i) at (2,1.8-\i*0.9) {};
  }
  \node[above=0.2cm of H23] {\scriptsize Linear + Tanh};

  \node[circle,draw,minimum size=1cm,fill=red!10] (O1) at (3,0) {$g(r)$};
  \node[below=0.3cm of O1] {\scriptsize Linear};

  \foreach \i in {1,...,5} {
    \draw[->,thick] (I1) -- (H1\i);
  }

  \foreach \i in {1,...,5} {
    \foreach \j in {1,...,5} {
      \draw[->,opacity=0.3] (H1\i) -- (H2\j);
    }
  }

  \foreach \i in {1,...,5} {
    \draw[->,thick] (H2\i) -- (O1);
  }
\end{tikzpicture}
\end{center}
\caption{Architecture of the ANN used to learn the profile $V(r)$ by learning function the $g(r)$ and using equation (\ref{Vg})}
\label{fig:ANN3}
\end{figure}
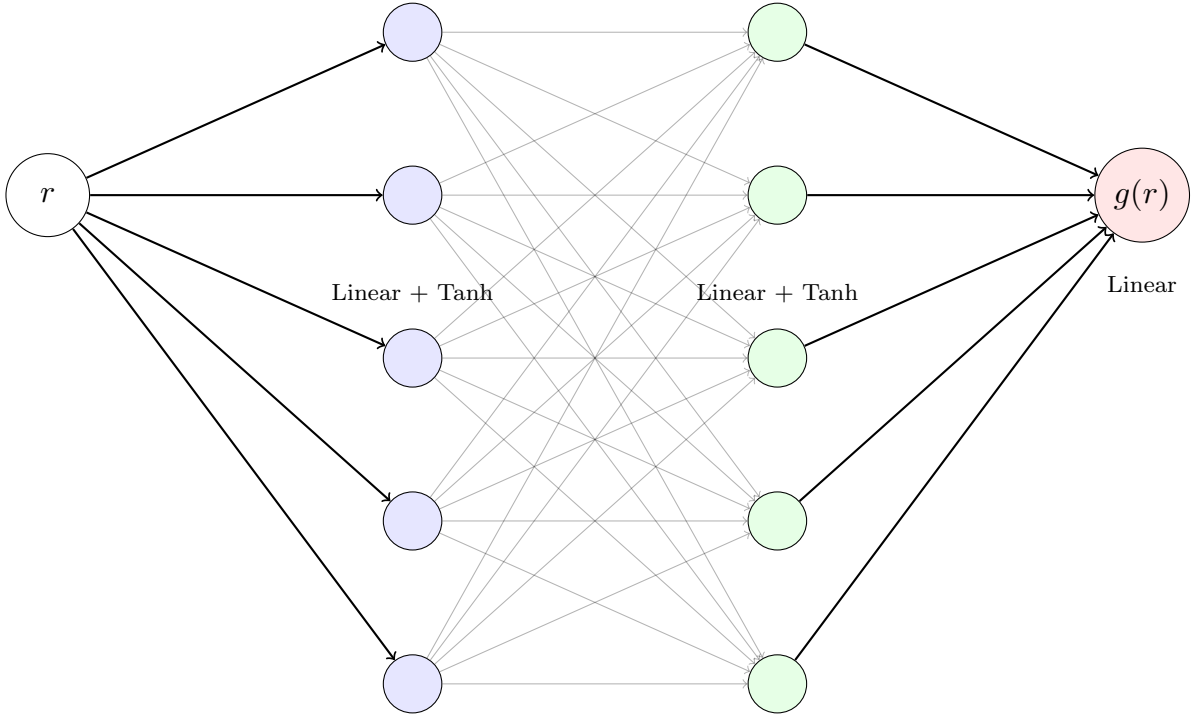

Here we will focus on learning the geometry when the background has a non-extremal horizon at some finite $r_0$, this implies that both $f_1(r)$ and $V(r)$ have a simple zero at $r = r_0$. Next we parametrize the potential $V(r)$ with an artificial neural network with the same architecture as in figure~\ref{fig:ANN1} but with different physical constraints. See figure~\ref{fig:ANN3}.

The parameter $r$ is in the interval $[r_0, r_{max}]$ and the physical constraints that $V(r_0) = 0$ at the horizon and $V(r_{max}) = 1$ is imposed via:
\begin{equation}\label{Vg}
V(r) = \frac{g(r) - g(r_0)}{g(r_{max}) - g(r_0)}\ .
\end{equation}
To solve the inverse problem—that is, to simultaneously learn the D7-brane embedding $L(\rho,m)$ and the unknown radial potential $V(r)$ from observed free energy data—we adopt a joint optimization strategy. A naive approach might involve a nested training loop: an outer loop updating the parameters of $V(r)$, and an inner loop, for each new $V(r)$, finding the optimal $L(\rho,m)$ by minimizing the DBI action. However, this nested approach is computationally highly inefficient.

Instead, we employ an \textbf{alternating optimization scheme} that trains both the network parameterizing $L(\rho,m)$ (denoted as $L_{\text{model}}$) and the network parameterizing $V(r)$ (denoted as $V_{\text{model}}$) concurrently. This method relies on two distinct loss functions:

\begin{enumerate}
    \item \textbf{Data-driven Loss ($\mathcal{L}_V$):} This loss quantifies how well the free energy $F(m)$ (computed from the current $L_{\text{model}}$ and $V_{\text{model}}$ for a given mass $m$) matches the empirically given free energy data $F_{\text{given}}(m)$. We define this as the mean squared error:
    $$ \mathcal{L}_V = \frac{1}{N_m} \sum_{i=1}^{N_m} \left( F\left(m^{(i)}\right) - F_{\text{given}}\left(m^{(i)}\right) \right)^2 $$
    where $N_m$ is the mini-batch size for masses, and $m^{(i)}$ are sampled masses. This loss function drives the $V_{\text{model}}$ to reconstruct a potential that yields the observed free energy curve.

    \item \textbf{Physical Loss ($\mathcal{L}_L$):} This loss is simply the regularized DBI action, $I_{D7}$, given by equation (\ref{I7_V}). This term ensures that for the current potential $V(r)$, the learned embedding $L(\rho,m)$ minimizes the action, thus satisfying the physical equations of motion.
    $$ \mathcal{L}_L = \frac{1}{N_m} \sum_{i=1}^{N_m} I_{D7}\left(L_{\text{model}}(\rho, m^{(i)}), V_{\text{model}}(r)\right) $$
\end{enumerate}

The training proceeds iteratively. At each optimization step, we sample a mini-batch of masses. We then alternate between updating the two networks:
\begin{itemize}
    \item In ``even'' steps, we compute $\mathcal{L}_V$ and use its gradients to update only the parameters of the $V_{\text{model}}$ network. This step adjusts the potential to better match the external free energy data.
    \item In ``odd'' steps, we compute $\mathcal{L}_L$ and use its gradients to update only the parameters of the $L_{\text{model}}$ network. This step adjusts the embedding profile to minimize the DBI action for the current (and recently updated) potential.
\end{itemize}
This alternating strategy (implemented in the \texttt{train\_inverse} function of the accompanying codebase\footnote{Available at \href{https://github.com/vesofilev/pinn_d7}{https://github.com/vesofilev/pinn\_d7}}) efficiently decouples the updates while ensuring both the data consistency for the potential and the physical validity of the embeddings. PyTorch's automatic differentiation engine handles the computation of gradients for both loss functions with respect to their respective network parameters. This allows us to learn a self-consistent geometry $V(r)$ and its corresponding D7-brane embeddings $L(\rho,m)$.

To apply this method we use ODE data for the meson melting set-up obtained using Mathematica as $F_{\text{given}}(m)$ and learn the ANNs for $V(r)$ and $L(\rho, m)$ using the alternating optimization scheme described above.  As one can see from equation (\ref{I7hot}) the potential that we would like to recover is:
\begin{equation}\label{pot_hot}
V(r) = 1 - \frac{1}{16 r^8}\ ,
\end{equation}
where the factor of 16 in the denominator is related to the choice of scale $r_0 = 1/\sqrt{2}$ for the position of the horizon in the radial variable $r$, which corresponds to choosing $u_0 = 1$ in the metric (\ref{BH-u}).

\begin{figure}
\hspace{-1cm}
\includegraphics[scale=0.7]{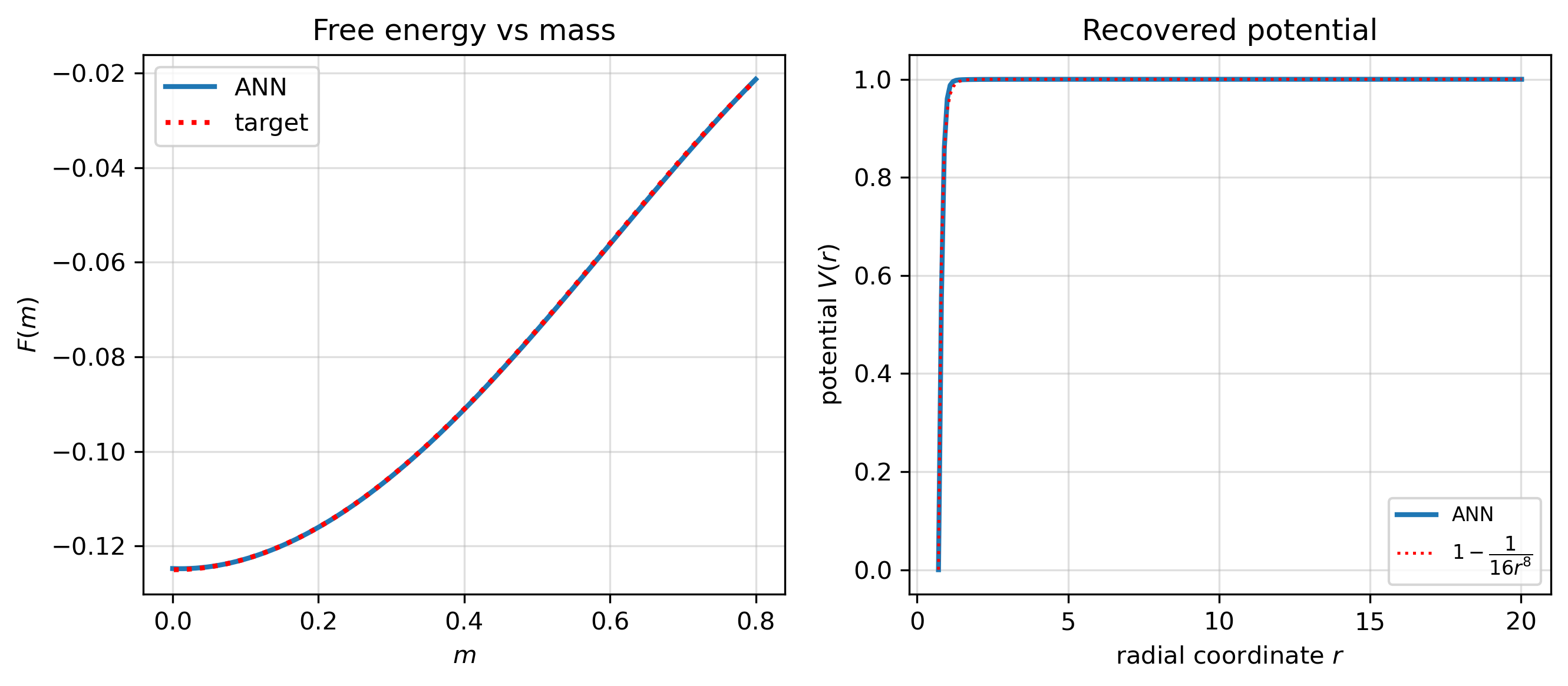}
\caption{Comparison of the target data and the learned data for the free energy versus mass curve as well as between the learned potential $V(r)$ and the potential used to generate the data.}
\label{fig:Learned}
\end{figure}

In figure~\ref{fig:Learned} we have presented plots of the learned free energy versus mass curves and the potential $V(r)$ (obtained after training for 500000 epochs). As one can see the optimization scheme perfectly matches the ODE data for the free energy and at the same times learns/recovers the steep potential (\ref{pot_hot}). In the spirit of reference~\cite{Hashimoto:2018ftp} this can be regarded as reconstructing the geometry of the AdS-black hole based on the field theory data (the free energy data $F_{\text{given}}(m)$).

\section{Comparison of ODE solver and Neural networks}
While the main advantage of the proposed ANN-based method is for inverse problems, it is still useful to compare quantitatively the computational effort between a mainstream numerical ODE solver and the ANN approach. Our metric is the total CPU time needed to obtain the free energy versus bare mass curve shown in figure~\ref{fig:F-mag-condit} using the conditional neural network. Before comparing runtimes, we must specify the number of epochs used to train the network. To this end, we use high-precision ODE data for the free energy versus bare mass to estimate the mean absolute error (MAE) of the ANN free energies on the same mass grid. We do the same for the condensate versus bare mass curves and plot how the MAE depends on the number of epochs. The idea is to identify the minimum number of epochs at which the MAE is close to saturation and use this epoch count to compare CPU times across methods.

\begin{figure}
\includegraphics[scale=0.7]{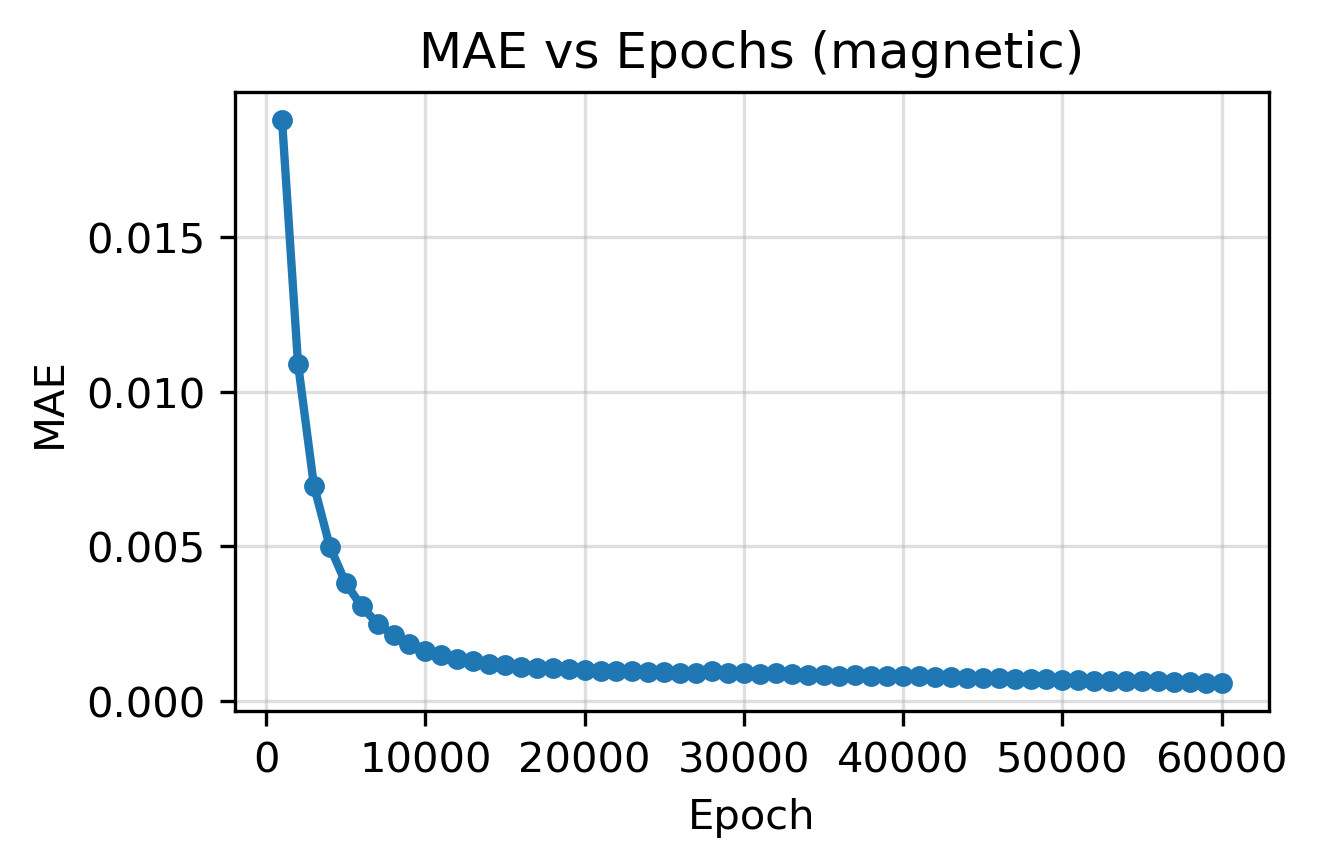}
\includegraphics[scale=0.7]{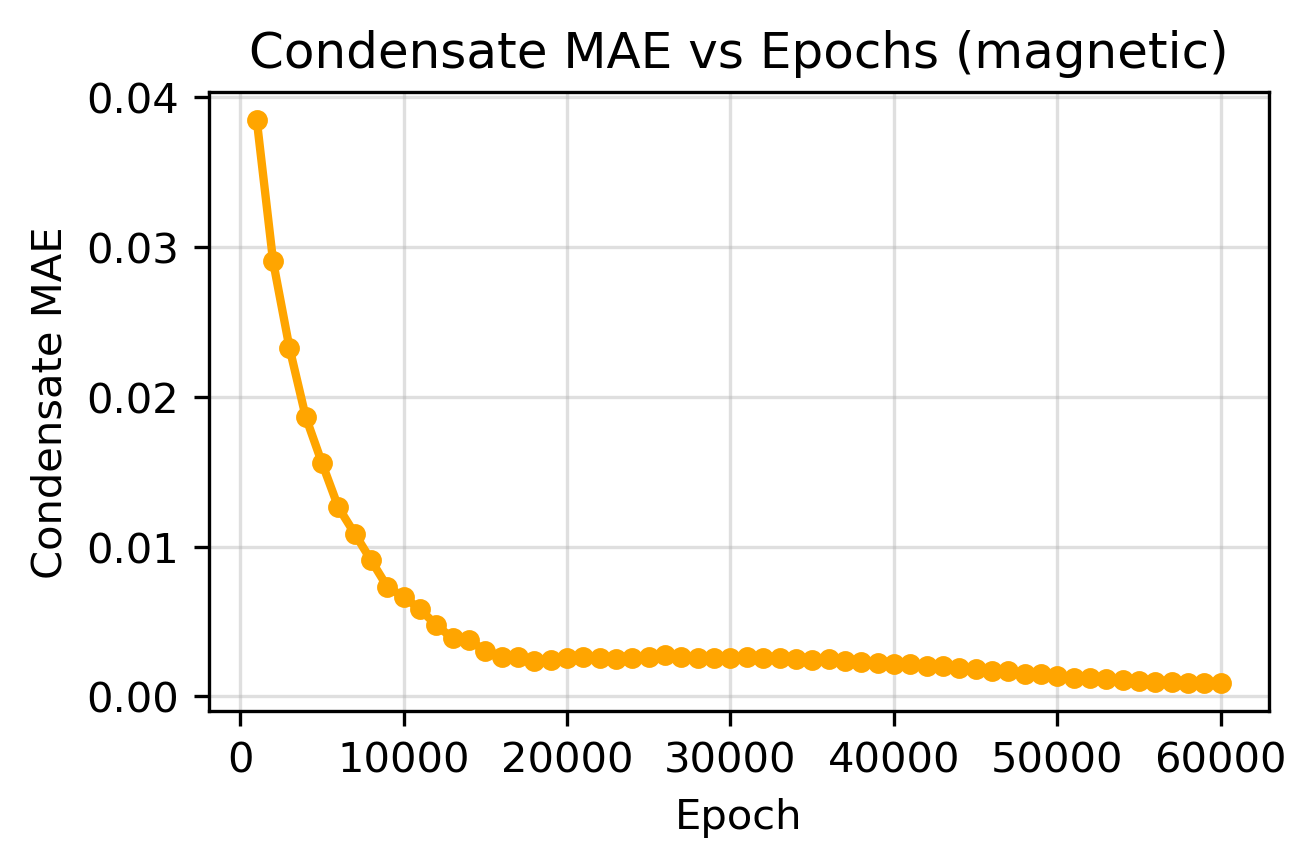}
\caption{Plots of the MAE for the free energy and condensate vs bare mass curves. }
\label{fig:MAE-FC}
\end{figure}

In figure~\ref{fig:MAE-FC} we present the MAE for the free energy and condensate curves. As one can see, for 20,000 epochs both MAE curves enter a steady “saturation” phase; therefore, we evolve the conditional network for 20,000 epochs to measure the total CPU time.

In table~\ref{tab:cpu_times}, we present the CPU times for solving the ODE using Python’s SciPy library, Mathematica, and our conditional ANN.

\begin{table}[h!]
\centering
\begin{tabular}{l c c}
\hline
\textbf{Method} & \textbf{CPU time (seconds)} & comp. factor\\
\hline
ODE scipy       & 1.32 &  - \\
ODE Mathematica & 33 & 25.0   \\
ANN             & 42 & 31.8  \\
\hline
\end{tabular}
\caption{Comparison of CPU times for different methods. The ‘comp. factor’ column reports time relative to the SciPy ODE baseline.}
\label{tab:cpu_times}
\end{table}

According to this measure, the ANN method is 31.8 times more expensive than the ODE method using specialized Python libraries. However, this comparison is valid only when the goal is to extract physically correct results with limited accuracy. As figure~\ref{fig:MAE-FC} suggests, convergence of the MAE in the “saturation” phase is slow and the best achievable accuracy is more limited. For example, if we want a relative error of $10^{-4}$ for the free energy, we need at least 100k epochs—an additional factor of 5 in runtime—making the ANN about 159 times slower than SciPy in that regime. It is also fair to note that the ODE approach has limitations, especially for stiff ODEs or when IR boundary conditions are delicate.

In conclusion, the main advantage of the ANN method is in solving inverse problems. For useful forward results at moderate accuracy, the ANN CPU time is comparable to Mathematica. For highly accurate results, the ANN converges slowly and is of order $10^2$ slower than the SciPy-based ODE solver when pushed to its optimal error.

\section{Conclusion}
In this work, we have shown that physics‐informed artificial neural networks can be applied efficiently to compute probe‐brane embeddings in holography via use of the regularized Dirac–Born–Infeld (DBI) action as a differentiable loss function. This approach not only produces the embedding profiles directly, but the loss function itself provides a measure of the system’s free energy. Remarkably, without solving the equations of motion explicitly or using derivatives higher than first order, our method can:
\begin{itemize}
\item reproduces the results for magnetic catalysis in the D3/D7 system. 
\item is sufficiently precise to capture the first order meson-melting transition including both Minkowski and black-hole embeddings. 
\item allows the efficient training of a single ``conditional'' network $L(\rho, m)$ which learns the entire one-parameter family of embeddings and allows efficient computation of first derivative of the free energy. 
\item solves the inverse problem by jointly training the embeddings ANN and the potential network $V(r)$ using two loss functions one ensuring the provided field theory data is matched and the other extremizing the DBI action.  This approach allows the reconstruction of the gravity dual geometry in the spirit of ref.~\cite{Hashimoto:2018ftp} which we demonstrated for the AdS–Schwarzschild geometry.
\end{itemize}

The ANNs employed in our study operate well within an under-parameterized regime. The single-mass and conditional models use shallow multilayer perceptrons (MLPs) with two hidden layers and 10–20 neurons per layer, corresponding to $O(10^2–10^3)$ trainable parameters (e.g., $\approx 140–480$ weights for 1D/2D inputs, respectively). In contrast, each optimization step evaluates the regularized DBI functional on a dense nonuniform $\rho$-grid with $N \approx 2000$ collocation points (and, in the conditional and inverse setups, across sampled masses), so the number of effective constraints per step exceeds the number of parameters by more than an order of magnitude.  Hard boundary conditions are enforced algebraically—$L(\rho_{\rm max})=m$ by construction and $V(r_0)=0$,  $V(r_{\rm max})=1$ via a linear rescaling--which further reduces the functional freedom. Empirically, we observe no overfitting: the conditional network generalizes across unseen masses and accurately reproduces both $F(m)$ and $\partial F/\partial m$, while discrepancies near the first-order transition are attributable to the coexistence of physical branches rather than model capacity. These considerations establish that the learning dynamics are governed by the physics-driven loss and not by excess expressivity of the networks.

Our results demonstrate several advantages over the numerical shooting methods: one does not need to investigate the IR boundary conditions. This is particularly useful when solving the inverse problem, since the equations of motion can develop stiff singularities as the potential varies. In fact, our attempt to solve the inverse problem by using neural differential equations to solve the equations of motion was met with this challenge. 

We also benchmarked runtime against conventional ODE shooting. For the magnetic setup at the “saturation” accuracy reached in 20k epochs, the conditional ANN takes $\approx 42$ seconds of CPU time, compared to $\approx 1.32$ seconds for a SciPy ODE solver and $\approx 33$  seconds for Mathematica --  about 32 times slower than SciPy but comparable to Mathematica. Pushing to higher accuracy ($10^{-4}$) requires at least 100k epochs, making the ANN roughly $10^2$ times slower than SciPy and about $10^1$ times slower than Mathematica. These results underscore that the ANN’s main advantage is not raw speed on forward problems but robustness and flexibility for inverse tasks and branch tracking, where deriving or integrating stiff EOMs is impractical.

There are several possible extensions of our work that we would like to pursue in the future:
\begin{itemize}
\item Extend to other probe‐brane systems (D(p)/D(q)), including nontrivial worldvolume gauge fields (chemical potentials, electric fields) and backreacted/flavored Veneziano–limit geometries.
\item Incorporate additional physical data (e.g. meson spectrum) into the inverse‐geometry program to reconstruct full metric functions ($f_1(r)$, $f_2(r)$).
\item Explore more sophisticated ANN architectures to improve convergence of the training process and handle more complex set-ups.
\end{itemize}

\section{Acknowledgments}
I would like to thank D. O'Connor for useful discussions. This work was supported by the Bulgarian NSF grant KP-06-N88/1.


\newpage

\begin{thebibliography}{99}

\bibitem{Maldacena:1997re}
J.~M.~Maldacena,
``The Large $N$ limit of superconformal field theories and supergravity,''
Adv. Theor. Math. Phys. \textbf{2}, 231-252 (1998)
doi:10.4310/ATMP.1998.v2.n2.a1
[arXiv:hep-th/9711200 [hep-th]].

\bibitem{Karch:2002sh}
A.~Karch and E.~Katz,
``Adding flavor to AdS / CFT,''
JHEP \textbf{06}, 043 (2002)
doi:10.1088/1126-6708/2002/06/043
[arXiv:hep-th/0205236 [hep-th]].


\bibitem{Filev:2007gb}
V.~G.~Filev, C.~V.~Johnson, R.~C.~Rashkov and K.~S.~Viswanathan,
``Flavoured large N gauge theory in an external magnetic field,''
JHEP \textbf{10}, 019 (2007)
doi:10.1088/1126-6708/2007/10/019
[arXiv:hep-th/0701001 [hep-th]].

\bibitem{Karch:2005ms}
A.~Karch, A.~O'Bannon and K.~Skenderis,
``Holographic renormalization of probe D-branes in AdS/CFT,''
JHEP \textbf{04}, 015 (2006)
doi:10.1088/1126-6708/2006/04/015
[arXiv:hep-th/0512125 [hep-th]].


\bibitem{Babington:2003vm}
J.~Babington, J.~Erdmenger, N.~J.~Evans, Z.~Guralnik and I.~Kirsch,
``Chiral symmetry breaking and pions in nonsupersymmetric gauge / gravity duals,''
Phys. Rev. D \textbf{69}, 066007 (2004)
doi:10.1103/PhysRevD.69.066007
[arXiv:hep-th/0306018 [hep-th]].

\bibitem{Mateos:2006nu}
D.~Mateos, R.~C.~Myers and R.~M.~Thomson,
``Holographic phase transitions with fundamental matter,''
Phys. Rev. Lett. \textbf{97}, 091601 (2006)
doi:10.1103/PhysRevLett.97.091601
[arXiv:hep-th/0605046 [hep-th]].


\bibitem{Albash:2006ew}
T.~Albash, V.~G.~Filev, C.~V.~Johnson and A.~Kundu,
``A Topology-changing phase transition and the dynamics of flavour,''
Phys. Rev. D \textbf{77}, 066004 (2008)
doi:10.1103/PhysRevD.77.066004
[arXiv:hep-th/0605088 [hep-th]].

\bibitem{Hashimoto:2018ftp}
K.~Hashimoto, S.~Sugishita, A.~Tanaka and A.~Tomiya,
``Deep learning and the AdS/CFT correspondence,''
Phys. Rev. D \textbf{98}, no.4, 046019 (2018)
doi:10.1103/PhysRevD.98.046019
[arXiv:1802.08313 [hep-th]].

\bibitem{Ahn:2024jkk}
B.~Ahn, H.~S.~Jeong, K.~Y.~Kim and K.~Yun,
``Holographic reconstruction of black hole spacetime: machine learning and entanglement entropy,''
JHEP \textbf{01}, 025 (2025)
doi:10.1007/JHEP01(2025)025
[arXiv:2406.07395 [hep-th]].

\bibitem{Ahn:2024gjf}
B.~Ahn, H.~S.~Jeong, K.~Y.~Kim and K.~Yun,
``Deep learning bulk spacetime from boundary optical conductivity,''
JHEP \textbf{03}, 141 (2024)
doi:10.1007/JHEP03(2024)141
[arXiv:2401.00939 [hep-th]].

\bibitem{Lei:2025loy}
M.~Lei and C.~Baehr,
``Geometric Meta-Learning via Coupled Ricci Flow: Unifying Knowledge Representation and Quantum Entanglement,''
[arXiv:2503.19867 [cs.LG]].

\bibitem{Halverson:2024axc}
J.~Halverson, J.~Naskar and J.~Tian,
``Conformal Fields from Neural Networks,''
[arXiv:2409.12222 [hep-th]].

\bibitem{Hashimoto:2024aga}
K.~Hashimoto, Y.~Hirono, J.~Maeda and J.~Totsuka-Yoshinaka,
``Neural network representation of quantum systems,''
Mach. Learn. Sci. Tech. \textbf{5}, no.4, 045039 (2024)
doi:10.1088/2632-2153/ad81ac
[arXiv:2403.11420 [hep-th]].

\bibitem{Mansouri:2024uwc}
M.~Mansouri, K.~Bitaghsir Fadafan and X.~Chen,
``Holographic complex potential of a quarkonium from deep learning,''
[arXiv:2406.06285 [hep-ph]].

\bibitem{Chen:2024ckb}
X.~Chen and M.~Huang,
``Machine learning holographic black hole from lattice QCD equation of state,''
Phys. Rev. D \textbf{109}, no.5, L051902 (2024)
doi:10.1103/PhysRevD.109.L051902
[arXiv:2401.06417 [hep-ph]].

\bibitem{Kou:2025qsg}
W.~Kou, X.~Lin, B.~Guo and X.~Chen,
``Physics-Informed Neural Network Approach to Quark-Antiquark Color Flux Tube,''
[arXiv:2506.03513 [hep-ph]].

\bibitem{Takayanagi:2025ula}
T.~Takayanagi,
``Essay: Emergent Holographic Spacetime from Quantum Information,''
Phys. Rev. Lett. \textbf{134}, no.24, 240001 (2025)
doi:10.1103/pg4r-fy8n
[arXiv:2506.06595 [hep-th]].

\bibitem{Sahay:2024vfw}
R.~Sahay, M.~D.~Lukin and J.~Cotler,
``Emergent Holographic Forces from Tensor Networks and Criticality,''
Phys. Rev. X \textbf{15}, no.2, 021078 (2025)
doi:10.1103/PhysRevX.15.021078
[arXiv:2401.13595 [quant-ph]].

\bibitem{Akers:2024wre}
C.~Akers, A.~Bouland, L.~Chen, T.~Kohler, T.~Metger and U.~Vazirani,
``Holographic pseudoentanglement and the complexity of the AdS/CFT dictionary,''
[arXiv:2411.04978 [hep-th]].

\bibitem{Park:2023slm}
C.~Park, S.~Kim and J.~H.~Lee,
``Holography Transformer,''
[arXiv:2311.01724 [hep-th]].

\bibitem{Paszke:2019}
A.~Paszke, S.~Gross, F.~Massa, A.~Lerer, J.~Bradbury, G.~Chanan, T.~Killeen, Z.~Lin, N.~Gimelshein, L.~Antiga, A.~Desmaison, A.~Kopf, E.~Yang, Z.~DeVito, M.~Raison, A.~Tejani, S.~Chilamkurthy, B.~Steiner, L.~Fang, J.~Bai and S.~Chintala,
``PyTorch: An Imperative Style, High‑Performance Deep Learning Library,''
Advances in Neural Information Processing Systems 32 (NeurIPS 2019),
arXiv:1912.01703. doi:10.48550/arXiv.1912.01703. (papers.nips.cc)

\bibitem{Kingma:2014adam}
D.~P.~Kingma and J.~Ba,
“Adam: A Method for Stochastic Optimization,”
arXiv:1412.6980 [cs.LG] (2014).
doi:10.48550/arXiv.1412.6980.
Published as a conference paper at ICLR 2015. (arxiv.org)

\end{thebibliography}
\end{document}